\documentclass[aps]{revtex4}
\usepackage{eurosym}
\usepackage{amsfonts}
\usepackage{amsmath}
\usepackage{amssymb,epsf}
\usepackage{color}
\usepackage{xcolor}
\usepackage{graphicx}
\usepackage{epstopdf}
\usepackage{float}
\usepackage{caption}
\usepackage{subfig}
\usepackage{hyperref}

\begin{document}

\title{Super-entropic black holes in gravity's rainbow and determining
constraints on rainbow functions}
\author{Behzad Eslam Panah$^{1}$\footnote{
email address: eslampanah@umz.ac.ir}, Narges Heidari$^{1}$, Mana Soleimani$%
^{1}$, and Maryam Kaveh$^{1}$}
\affiliation{$^{1}$ Department of Theoretical Physics, Faculty of Basic Sciences,
University of Mazandaran, P. O. Box 47416-95447, Babolsar, Iran}

\begin{abstract}
This paper is motivated by the application of the inverse isoperimetric
inequality to establish constraints on the parameters of gravity's rainbow.
We investigate the thermodynamic (in)stability conditions for $d-$%
dimensional energy-dependent black holes, which are recognized as $d-$%
dimensional black holes within the framework of gravity's rainbow. To
achieve this, we calculate thermodynamic quantities such as Hawking
temperature, entropy, total mass, and heat capacity in both extended and
non-extended phase spaces for these black holes. We assess the physical and
stable regions by utilizing these thermodynamic quantities alongside the
inverse isoperimetric inequality, aiming to determine constraints on the
rainbow functions. Finally, we show that by considering a constraint on the
rainbow function, these black holes satisfy the super-entropic condition.
\end{abstract}

\maketitle

\section{Introduction}

Black holes exert a strong gravitational force, making it essential to
explore alternative theories of gravity in high-energy physics, particularly
when considering particle energies approaching the Planck energy $E_{p}$.
One such alternative is gravity's rainbow, a theory that generalizes doubly
special relativity to incorporate curvature, as introduced by Magueijo and
Smolin \cite{Magueijo2004}. In this framework, spacetime is influenced by
energy functions, which indicate that particles interacting with spacetime
can access specific energy ranges. This results in the emergence of a
rainbow of energy levels.

Due to the high-energy properties of black holes, it is essential to
consider quantum corrections to classical perspectives. One approach to
incorporate these quantum corrections into gravitational fields is by
utilizing energy-dependent spacetime. Research has demonstrated that the
quantum corrections in gravitational systems can be observed as a function
of the energy of the particles probing the spacetime, which aligns with the
concept of gravity's rainbow \cite{Smolin2006,Ling2007,GarattiniM2012}. By
adopting this energy-dependent perspective, we anticipate observing its
effects on black hole properties, particularly concerning their
thermodynamics. This serves as the primary motivation for exploring the
generalization of gravity's rainbow.

Additionally, there have been specific advancements related to gravity's
rainbow in the context of black holes. Notable examples include
modifications to the energy-momentum dispersion relation, which are
supported by studies in string theory \cite{Kostelecky1989}, loop quantum
gravity \cite{Gambini1999}, and experimental observations \cite{Abraham2010}%
. The possibility of black hole remnants \cite{Ali2015} has been suggested
as a potential solution to the information paradox \cite{Ali2015b}. Other
contributions include the development of a modified uncertainty principle 
\cite{Li2009}, establishing a connection to Horava-Lifshitz gravity \cite%
{GarattiniS2015}, and providing a UV completion of Einstein gravity \cite%
{Magueijo2004}.

Furthermore, in the realm of cosmology, this theory of gravity offers a
possible resolution to the Big Bang singularity problem \cite%
{BigB1,BigB2,BigB3}. In astrophysics, it also addresses the existence of
compact objects with masses greater than $2.5M_{\odot}$, which can fill the
mass-gap regime \cite{MassCom1,MassCom2}. These motivations have led us to
consider gravity's rainbow in our study, particularly focusing on the
thermodynamic properties of black holes.

There are three known models for the rainbow functions 
\begin{equation}
\left\{ 
\begin{array}{ccc}
f\left( \varepsilon \right) =\frac{e^{\beta \varepsilon }-1}{\beta
\varepsilon },~~\&~~g\left( \varepsilon \right) =1, &  & \text{model (i)} \\ 
&  &  \\ 
f\left( \varepsilon \right) =1,~~\&~~g\left( \varepsilon \right) =\sqrt{%
1-\eta \varepsilon ^{n}}, &  & \text{model (ii)} \\ 
&  &  \\ 
f\left( \varepsilon \right) =g\left( \varepsilon \right) =\frac{1}{1-\lambda
\varepsilon }, &  & \text{model (iii)}%
\end{array}%
\right. ,  \label{Model}
\end{equation}%
where $\beta $, $\eta $, and $\lambda $ are constants that can be adjusted
through experimentation. Additionally, $\varepsilon =\frac{E}{E_{P}}$, where 
$E$ and $E_{P}$ represent the particle and Planck energies, respectively.
This relationship imposes an upper limit on $\varepsilon $, such that $%
\varepsilon \leq 1$. Model (i) is associated with the hard spectra observed
in gamma-ray bursts \cite{Amelino1998}. Model (ii) draws motivation from
research in loop quantum gravity and noncommutative geometry \cite%
{Jacob2010, Amelino2013}. Model (iii) arises from the principle of the
constancy of the speed of light \cite{Magueijo2002}.

After Hawking and Bekenstein introduced the connections between the
geometric properties of black holes and thermodynamic variables \cite%
{Bekenstein1973,Hawking1974}, the thermodynamics of black holes garnered
significant attention. This field has provided deep insights into the
relationship between the physical properties of gravity and classical
thermodynamics. Among the thermodynamic properties of black holes, phase
transitions and thermal stability have been extensively studied in the
literature \cite{Th1,Th2,Th3}.

Several methods exist for investigating the phase transitions of black
holes. In the context of the canonical ensemble, the heat capacity provides
insights into thermodynamic stability and phase transitions. Specifically,
the sign of the heat capacity indicates whether black holes are stable or
unstable. Black holes with a positive heat capacity are considered thermally
stable, while those with a negative heat capacity are deemed unstable.
Additionally, the divergence points of the heat capacity signify the phase
transitions between stable and unstable states. Furthermore, negative
Helmholtz free energy indicates the global stability of a thermodynamic
system.

Recently, there has been a growing interest in studying the thermodynamic
phase transitions of black holes in anti-de Sitter (AdS) spacetime within an
extended phase space, where the cosmological constant is treated as a
variable \cite{Thext1, Thext2}. This interest is primarily driven by the
AdS/CFT duality \cite{AdSCFT}. One important quantity that provides insights
into the thermodynamic system is the reverse isoperimetric inequality. This
inequality indicates that a Schwarzschild black hole has the maximum entropy
for a given thermodynamic volume $V$. However, several black hole solutions,
known as super-entropic black holes, violate this inequality \cite%
{Klemm2014,Hennigar2015,Hennigar2015b,Brenna2015,Sinamuli2016,Noorbakhsh2017,Feng2017,WuWYW2020,WuWYW2020b,Xu2020,WuWWY2021,WuWu2021,Wu2021,Wu2021b,Boudet2021,Zhang2021,Dehghani2023}%
. Recent investigations have focused on the thermodynamic stability of these
solutions, leading to the conjecture that super-entropic black holes are
thermodynamically unstable \cite%
{Cong2019,Johnson2020,EslamPanah2025,JohnsonarXiv}.

Our primary goal is to establish constraints on rainbow functions by
utilizing the inverse isoperimetric inequality and thermodynamic stability
conditions. In this context, we will identify which of these rainbow
functions are appropriate for studying stable black holes from a
thermodynamic perspective.

The action of gravity's rainbow in the presence of the cosmological constant
is given by \cite{Magueijo2004} 
\begin{equation}
\mathcal{I}=\frac{1}{2\kappa ^{2}}\int_{\partial \mathcal{M}}d^{d}x\sqrt{-g}%
\left[ R(\varepsilon )-2\Lambda (\varepsilon )\right] ,  \label{action}
\end{equation}%
where $\kappa ^{2}=8\pi G(\varepsilon )$, with $G(\varepsilon )$
representing an energy-dependent Newton's constant. Additionally, $%
R(\varepsilon )$ and $\Lambda (\varepsilon )$ denote the energy-dependent
Ricci scalar and the energy-dependent cosmological constant, respectively.
Furthermore, $g=\det (g_{\mu \nu })$ is the determinant of the metric tensor 
$g_{\mu \nu }$. Hereafter, we will assume that $G(\varepsilon )=1$.

By varying the action (\ref{action}) with respect to the metric tensor $%
g_{\mu \nu }$, the gravitational field equations is given by \cite%
{Magueijo2004} 
\begin{equation}
G(\varepsilon)_{\mu \nu }+\Lambda(\varepsilon) g_{\mu \nu }=0,  \label{FE1}
\end{equation}%
where $G(\varepsilon)_{\mu \nu }$ is the energy-dependent Einstein's tensor.

\section{Thermodynamics stability conditions}

We consider a $d$-dimensional energy-dependent spacetime in the following
form 
\begin{equation}
ds^{2}=-\frac{\psi (r,\varepsilon )}{f^{2}\left( \varepsilon \right) }dt^{2}+%
\frac{\left( \frac{dr^{2}}{\psi (r,\varepsilon )}+r^{2}d\Omega ^{2}\right) }{%
g^{2}\left( \varepsilon \right) },  \label{Metric}
\end{equation}%
where 
\begin{equation}
d\Omega ^{2}=d\theta _{1}+\sum_{i=2}^{d-2}\prod\limits_{j=1}^{i-1}\sin
^{2}\theta _{j}d\theta _{i}^{2},
\end{equation}%
and $\psi (r,\varepsilon )$ is the metric function. In the above
energy-dependent spacetime, $f\left( \varepsilon \right) $ and $%
g\left(\varepsilon \right) $ are known rainbow functions defined in Eq. (\ref%
{Model}).

The metric function is obtained by using the gravitational field equations (%
\ref{FE1}) and energy-dependent spacetime (\ref{Metric}) in the following
form \cite{Hendi2017k} 
\begin{equation}
\psi (r,\varepsilon )=1-\frac{m_{0}}{r^{d-3}}-\frac{2\Lambda \left(
\varepsilon \right) r^{2}}{\left( d-1\right) \left( d-2\right) g^{2}\left(
\varepsilon \right) }.  \label{f(r)}
\end{equation}%
where $m_{0}$ is a constant which is related to the mass of black hole.

In order to investigate the thermodynamic properties of the $d-$dimensional
black hole solutions in gravity's rainbow, it is crucial to establish the
relationship between the mass ($m_{0}$), the event horizon radius ($r_{+}$),
and the cosmological constant. This can be achieved by setting $g_{tt}=\psi
(r)$ equal to zero (i.e. $g_{tt}=\psi (r,\varepsilon)=0$). So, we have 
\begin{equation}
m_{0}=-\frac{2\Lambda \left( \varepsilon \right) r_{+}^{d-1}}{\left(
d-1\right) \left( d-2\right) g^{2}\left( \varepsilon \right) }+r^{d-3}.
\label{mm}
\end{equation}

The surface gravity of the $d-$dimensional black holes in gravity's rainbow
is given by 
\begin{equation}
\kappa =\left. \frac{g\left( \varepsilon \right) \psi ^{\prime
}(r,\varepsilon )}{2f\left( \varepsilon \right) }\right\vert _{r=r_{+}},
\label{k0}
\end{equation}%
where $\psi ^{\prime }(r,\varepsilon )=\frac{d\psi (r,\varepsilon )}{dr}$.
Considering the metric function (\ref{f(r)}), and substituting the mass (\ref%
{mm}) into equation (\ref{k0}), we find the surface gravity for these black
holes as 
\begin{equation}
\kappa =\frac{\left( d-2\right) \left( d-3\right) g^{2}\left( \varepsilon
\right) -2\Lambda \left( \varepsilon \right) r_{+}^{2}}{2\left( d-2\right)
f\left( \varepsilon \right) g\left( \varepsilon \right) r_{+}}.  \label{kk}
\end{equation}%
Then, using the definition of the Hawking temperature $T=\frac{\kappa }{2\pi 
}$, we find it as \cite{Hendi2017k} 
\begin{equation}
T=\frac{\left( d-2\right) \left( d-3\right) g^{2}\left( \varepsilon \right)
-2\Lambda \left( \varepsilon \right) r_{+}^{2}}{4\left( d-2\right) \pi
f\left( \varepsilon \right) g\left( \varepsilon \right) r_{+}}.
\label{TemF(R)CPMI}
\end{equation}

To have positive temperature, we have to respect $\left( d-2\right) \left(
d-3\right) g^{2}\left( \varepsilon \right) -2\Lambda \left( \varepsilon
\right) r_{+}^{2}>0$, which leads to a constraint on the rainbow function $%
g^{2}\left( \varepsilon \right) $ as 
\begin{equation}
g^{2}\left( \varepsilon \right) >\frac{2\Lambda \left( \varepsilon \right)
r_{+}^{2}}{\left( d-2\right) \left( d-3\right) }.  \label{ConsT}
\end{equation}

Using the area law, the entropy of black holes in gravity's rainbow is given
by \cite{Hendi2017k} 
\begin{equation}
S=\frac{A}{4}=\frac{\omega _{d-2}}{4}\left( \frac{r_{+}}{g\left( \varepsilon
\right) }\right) ^{d-2},  \label{A}
\end{equation}%
where $A=\omega _{d-2}\left( \frac{r_{+}}{g\left( \varepsilon \right) }%
\right) ^{d-2}$ is the area of the event horizon. Also, $\omega_{d-2}$
stands for the area of the space orthogonal to constant ($t$, $r$) surfaces
and is defined as $\omega _{d-2}=\frac{2\pi ^{\left( \frac{d-1}{2}\right) }}{%
\Gamma \left( \frac{d-1}{2 }\right) }$.

By employing the Ashtekar-Magnon-Das (AMD) approach \cite{AMDI, AMDII}, the
total mass of the black hole is obtained as follows \cite{Hendi2017k} 
\begin{equation}
M=\frac{\left( d-2\right) \omega _{d-2}}{16\pi f\left( \varepsilon \right)
g^{d-3}\left( \varepsilon \right) }m_{0},  \label{AMDMass}
\end{equation}%
where by substituting the mass (\ref{mm}) within the equation (\ref{AMDMass}%
), yields 
\begin{equation}
M=\frac{\left( \left( d-2\right) g^{2}\left( \varepsilon \right) -\frac{%
2\Lambda \left( \varepsilon \right) r_{+}^{2}}{\left( d-1\right) }\right)
\omega _{d-2}r_{+}^{d-3}}{16\pi f\left( \varepsilon \right) g^{d-1}\left(
\varepsilon \right) },  \label{MM}
\end{equation}%
To ensure positive values for the total mass, we derive a constraint on the
rainbow function $g^{2}(\varepsilon)$, which is 
\begin{equation}
g^{2}\left( \varepsilon \right) >\frac{2\Lambda \left( \varepsilon \right)
r_{+}^{2}}{\left( d-1\right) \left( d-2\right) }.  \label{ConsM}
\end{equation}

\subsection{Local stability}

We evaluate the local stability of $d$-dimensional black holes in gravity's
rainbow by studying their heat capacity. In the context of the canonical
ensemble, a thermodynamic system's local stability can be examined through
its heat capacity. Discontinuities in heat capacity indicate potential
thermal phase transitions that the system may undergo. Furthermore, positive
heat capacity corresponds to thermal stability, while negative heat capacity
indicates instability. Additionally, the roots of the heat capacity are of
interest, as they may signal possible transitions between stable and
unstable states or bound points.

We find the heat capacity of $d-$dimensional black holes in gravity's
rainbow as 
\begin{equation}
C=\frac{T\left( \frac{\partial S}{\partial r_{+}}\right) }{\left( \frac{%
\partial T}{\partial r_{+}}\right) }=-\frac{\omega _{d-2}\left( \left(
d-3\right) g^{2}\left( \varepsilon \right) -\frac{2\Lambda \left(
\varepsilon \right) r_{+}^{2}}{\left( d-2\right) }\right) r_{+}^{d-2}}{%
4\left( \frac{\left( d-3\right) g^{2}\left( \varepsilon \right) }{\left(
d-2\right) }+\frac{2\Lambda \left( \varepsilon \right) r_{+}^{2}}{\left(
d-2\right) ^{2}}\right) g^{^{d-2}}\left( \varepsilon \right) }.
\end{equation}%
in order to satisfies the local stability in the canonical ensemble context
(i.e., $C>0$), we find two follwoing constrains 
\begin{equation}
\left\{ 
\begin{array}{c}
g^{2}\left( \varepsilon \right) >\frac{2\Lambda \left( \varepsilon \right)
r_{+}^{2}}{\left( d-2\right) \left( d-3\right) }, \\ 
\\ 
g^{2}\left( \varepsilon \right) <\frac{-2\Lambda \left( \varepsilon \right)
r_{+}^{2}}{\left( d-2\right) \left( d-3\right) },%
\end{array}%
\right. .  \label{ConstHeat}
\end{equation}%
whereas $g^{2}\left( \varepsilon \right) $ cannot be negative, the
energy-dependent cosmological constant must be negative (i.e., $\Lambda
\left( \varepsilon \right) <0$). This reveals that AdS black holes can
satisfy local stability conditions.

\subsection{Global stability}

Hawking and Page initially proposed the concept of investigating a black
hole's global stability \cite{Th2}. They suggested that this global
stability can be assessed within the framework of the grand canonical and
canonical ensembles by calculating the Gibbs and Helmholtz free energies,
respectively. Since we consider uncharged black holes, there is no
difference between the Gibbs and Helmholtz free energies. It is notable that
the negative of the Helmholtz free energy determines the global stability of
a thermodynamic system in the context of the canonical ensemble. In this
study, we aim to determine global stability using the Helmholtz free energy
for $d$-dimensional black holes in gravity's rainbow.

The Helmholtz free energy is defined by 
\begin{equation}
F=M-TS=\frac{\omega _{d-2}\left[ \left( d-1\right) g^{2}\left( \varepsilon
\right) +\frac{2\Lambda \left( \varepsilon \right) r_{+}^{2}}{\left(
d-2\right) }\right] r_{+}^{d-3}}{16\pi \left( d-1\right) f\left( \varepsilon
\right) g^{d-1}\left( \varepsilon \right) }.
\end{equation}

The global stability areas are defined when the Helmholtz free energy is
negative (i.e., $F<0$). To evaluate the global stability of the $d-$%
dimensional black hole in gravity's rainbow, we find a constraint on rainbow
function $g^{2}\left( \varepsilon \right) $ as 
\begin{equation}
g^{2}\left( \varepsilon \right) <\frac{-2\Lambda \left( \varepsilon \right)
r_{+}^{2}}{\left( d-1\right) \left( d-2\right) }.  \label{ConstHel}
\end{equation}

The constraint mentioned above indicates that AdS black holes can satisfy
the global stability condition, just as they do for local stability.
Consequently, the heat capacity and Helmholtz free energy of AdS black holes
simultaneously meet both local and global stability conditions.

\subsection{Extended Phase Space}

In extended phase space the effective cosmological constant plays the role
of a thermodynamic pressure, given by \cite{Thext1, Thext2} 
\begin{equation}
P=\frac{-\Lambda \left( \varepsilon \right) }{8\pi }.  \label{Lambda1}
\end{equation}

By substituting Eq. (\ref{A}) and Eq. (\ref{Lambda1}) into Eq. (\ref{MM}),
the total mass of the black hole can be expressed in terms of entropy and
thermodynamic 
\begin{equation}
M\left( S,P\right) =\frac{\omega _{d-2}\left( \frac{16\pi \left( \frac{4S}{%
\omega _{d-2}}\right) ^{\frac{d-1}{d-2}}P}{\left( d-1\right) }+\left(
d-2\right) \left( \frac{4S}{\omega _{d-2}}\right) ^{\frac{d-3}{d-2}}\right) 
}{16\pi f\left( \varepsilon \right) }.  \label{MSP}
\end{equation}

We can get the thermodynamic volume ($V$) of bumblebee AdS black holes by
using Eq. (\ref{MSP}), which leads to 
\begin{equation}
V=\left( \frac{\partial M\left( S,P\right) }{\partial P}\right) _{S}=\frac{%
\omega _{d-2}\left( \frac{4S}{\omega _{d-2}}\right) ^{^{\frac{d-1}{d-2}}}}{%
\left( d-1\right) f\left( \varepsilon \right) }=\frac{\omega _{d-2}}{\left(
d-1\right) f\left( \varepsilon \right) }\left( \frac{r_{+}}{g\left(
\varepsilon \right) }\right) ^{d-1}.  \label{V1}
\end{equation}

Using $M(S,P)$, we can express the temperature ($T$) in terms of
thermodynamic pressure ($P$) and entropy ($S$) in the following form 
\begin{equation}
T=\left. \frac{\partial M}{\partial S}\right\vert _{P}=\frac{\omega
_{d-2}\left( \frac{16\pi \left( \frac{4S}{\omega _{d-2}}\right) ^{\frac{d-1}{%
d-2}}P}{\left( d-2\right) }+\left( d-3\right) \left( \frac{4S}{\omega _{d-2}}%
\right) ^{\frac{d-3}{d-2}}\right) }{16\pi f\left( \varepsilon \right) S}.
\label{Text}
\end{equation}

It is notable that the heat capacity at constant volume ($C_{V}$) is zero
for static black holes in any modified theory of gravity when the
thermodynamic volume depends on entropy, i.e., $V\propto S$.

We get the heat capacity at constant pressure ($C_{P}$) as 
\begin{equation}
C_{P}=\frac{T}{\left. \frac{\partial T}{\partial S}\right\vert _{P}}=\frac{%
\left( d-2\right) \left( \frac{16\pi \left( \frac{4S}{\omega _{d-2}}\right)
^{\frac{2}{d-2}}P}{\left( d-2\right) }+\left( d-3\right) \right) S}{\frac{%
16\pi \left( \frac{4S}{\omega _{d-2}}\right) ^{\frac{2}{d-2}}P}{\left(
d-2\right) }-\left( d-3\right) }.  \label{CP1}
\end{equation}

To study the thermodynamic stability of $d-$dimensional AdS black holes in
gravity's rainbow within the extended phase space, we focus on the heat
capacity at constant pressure. The system satisfies the thermodynamic
stability condition when $C_{P}>0$, which imposes 
\begin{equation}
P>\frac{\left( d-2\right) \left( d-3\right) }{16\pi \left( \frac{4S}{\omega
_{d-2}}\right) ^{\frac{2}{d-2}}},  \label{ConP}
\end{equation}%
by substituting $P=\frac{-\Lambda(\varepsilon)}{8\pi}$ and $S = \frac{%
\omega_{d-2}}{4} \left( \frac{r_{+}}{g(\varepsilon)} \right)^{d-2}$ into Eq.
(\ref{ConP}), we find that the two constraints are similar to Eq. (\ref%
{ConstHeat}).

To achieve thermodynamic stability, we consider the positive values for the
Hawking temperature, total mass, local and global stability conditions, and
the heat capacity at constant pressure. These considerations lead to four
constraints (Eqs. (\ref{ConsT}), (\ref{ConsM}), (\ref{ConstHeat}), and (\ref%
{ConstHel})). By imposing the thermodynamic stability conditions and
evaluating the three conditions in Eqs. (\ref{ConsT}), (\ref{ConsM}), (\ref%
{ConstHeat}), and (\ref{ConstHel}), we can find a suitable range for the
rainbow function $g^{2}\left( \varepsilon \right)$ as 
\begin{equation}
\frac{2\Lambda \left( \varepsilon \right) r_{+}^{2}}{\left( d-1\right)
\left( d-2\right) }<g^{2}\left( \varepsilon \right) <\frac{-2\Lambda \left(
\varepsilon \right) r_{+}^{2}}{\left( d-1\right) \left( d-2\right) },
\label{Constraint}
\end{equation}%
where it indicates that the above constraint is valid for $d\geq 3$ and $%
\Lambda \left( \varepsilon \right) \neq 0$. In other words, by applying the
above constraint to $g^2(\varepsilon)$, the Hawking temperature, total mass,
and the heat capacity at constant pressure of the $d-$dimensional AdS black
holes in gravity's rainbow will be positive and satisfy thermodynamic
stability.

The constraint on $g^{2}(\varepsilon)$ (Eq. \ref{Constraint}) indicates that
it depends on the dimension of spacetime ($d$), the energy-dependent
cosmological constant ($\Lambda(\varepsilon)$), and the radius of black
holes ($r_{+}$). As a result, $g^{2}(\varepsilon)$ cannot be constant. This
finding implies that model (i) for the rainbow functions (i.e., $%
f(\varepsilon)=\frac{e^{\beta \varepsilon} - 1}{\beta \varepsilon}$ and $%
g(\varepsilon)=1$) is not appropriate for analyzing black holes as
thermodynamic systems, as these black holes do not meet the conditions for
thermodynamic stability.

It is straightforward to show that the conserved quantities and
thermodynamic quantities comply with the first law, in the following form 
\begin{equation}
dM=TdS+VdP.
\end{equation}

In addition, by considering Eqs. (\ref{MSP})-(\ref{Text}), the Smarr
relation is given by 
\begin{equation}
M=2(TS-PV).
\end{equation}

\section{\textbf{inverse isoperimetric inequality condition }}

In Refs. \cite{Cong2019,Johnson2020,EslamPanah2025,JohnsonarXiv}, a link
between super-entropic black holes and thermodynamic instability was
suggested. Confirming this connection would be of great importance. It is
important to note that super-entropic black holes arise from the violation
of the inverse isoperimetric inequality \cite%
{Klemm2014,Hennigar2015,Hennigar2015b,Brenna2015,Sinamuli2016,Noorbakhsh2017,Feng2017,Xu2020,Wu2021,Wu2021b,Boudet2021,Zhang2021,Dehghani2023}%
. A black hole is classified as a super-entropic black hole if it satisfies
the condition $\mathcal{R}<1$. This ratio is defined for $d-$dimensional
spacetime as \cite{Cvetic2011} 
\begin{equation}
\mathcal{R}=\left( \frac{\left( d-1\right) V}{\omega _{d-2}}\right) ^{\frac{1%
}{d-1}}\left( \frac{\omega _{d-2}}{A}\right) ^{\frac{1}{d-2}}=\frac{1}{f^{%
\frac{1}{d-1}}\left( \varepsilon \right) },  \label{R}
\end{equation}%
which indicates that $\mathcal{R}$ can be less than $1$ (i.e., $\mathcal{R}<1
$) for $f\left( \varepsilon \right) >1$. Therefore, the black holes in
gravity's rainbow are considered super-entropic black holes when the rainbow
function $f\left( \varepsilon \right) $ is greater than $1$ (i.e., $f\left(
\varepsilon \right) >1$). This implies that the black holes in gravity's
rainbow are thermodynamically unstable systems when $f\left( \varepsilon
\right) >1$.

On the other hand, the system achieves thermodynamic stability when $%
f\left(\varepsilon \right) \leq 1$, as this condition requires $\mathcal{R}
\geq 1$. Consequently, we cannot use model (iii), where $f\left(\varepsilon%
\right)=g\left(\varepsilon\right) = \frac{1}{1-\lambda\varepsilon}$, to
study black holes when $\lambda$ is positive, since these black holes
exhibit thermodynamic instability. Specifically, with $\lambda>0$ in the
model (iii), the rainbow function $f\left( \varepsilon \right)$ exceeds $1$
(i.e., $f\left( \varepsilon \right)>1$), which leads $\mathcal{R}<1$ and in
this case, the AdS black holes in gravity's rainbow are thermodynamic
unstable systems. However, for negative values of $\lambda$, we can use the
model (iii).

According to our analysis, model (ii) satisfies all thermodynamic stability
conditions without any limitations. Specifically, in model (ii), where $%
f\left( \varepsilon \right)=1$ and $g\left( \varepsilon \right) = \sqrt{1 -
\eta \varepsilon^{n}}$, both the conserved quantities and thermodynamic
quantities are positive, and $\mathcal{R} \geq 1$.

\section{\textbf{Conclusions}}

In this paper, we first considered the black hole solutions in $d-$%
dimensional energy-dependent spacetime. We investigated the thermodynamic
quantities to study thermodynamic stability in both extended and
non-extended phase spaces. Our analysis indicates that considering the
positivity of the Hawking temperature, total mass, heat capacity (or local
stability), the negative value of the Helmholtz free energy (global
stability) in non-extended phase space, and the positivity of the heat
capacity at constant pressure in extended phase space imposed a constraint
on $g^{2}\left(\varepsilon\right)$ in the form $\frac{2\Lambda \left(
\varepsilon \right) r_{+}^{2}}{\left( d-1\right)\left( d-2\right) }%
<g^{2}\left( \varepsilon \right) <\frac{-2\Lambda \left(\varepsilon \right)
r_{+}^{2}}{\left( d-1\right) \left( d-2\right) }$.

Based on the obtained constraint, $g^{2}\left( \varepsilon \right)$ could
not be constant. This implied that model (i), i.e., $f\left( \varepsilon
\right) =\frac{e^{\beta \varepsilon }-1}{\beta\varepsilon }$ and $g\left(
\varepsilon \right) =1$, was not suitable for studying black holes as
thermodynamic systems, as black holes modeled in this way could not satisfy
the conditions for thermodynamic stability.

Using the inverse isoperimetric inequality, we found that $d$-dimensional
AdS black holes in gravity's rainbow are classified as super-entropic black
holes when $f\left( \varepsilon \right) > 1$. Based on the relationship
between super-entropic black holes and thermodynamic instability, $d$%
-dimensional AdS black holes in gravity's rainbow exhibit thermodynamic
instability when $f\left( \varepsilon \right) > 1$. Conversely, for $f\left(
\varepsilon \right) \leq 1$, the inverse isoperimetric inequality yields $%
\mathcal{R} \geq 1$. In this case, we could not consider model (iii),
defined as $f\left( \varepsilon \right) =g\left( \varepsilon \right) =\frac{1%
}{1-\lambda\varepsilon}$, for studying black holes when $\lambda$ was
positive, because these black holes did not satisfy thermodynamic stability.
In other words, when considering $\lambda>0$ in the model (iii), $%
f\left(\varepsilon \right)>1$, which implies $\mathcal{R}<1$ and in this
case, the AdS black holes in gravity's rainbow were thermodynamic unstable
systems. However, for $\lambda<0$, the model (iii) could be suitable.

Our analysis revealed that model (ii) satisfies all thermodynamic stability
conditions without any limitations. Indeed, using model (ii), i.e., $%
f\left(\varepsilon \right) =1$ and $g\left( \varepsilon \right) =\sqrt{%
1-\eta \varepsilon ^{n}}$, the conserved quantities and thermodynamic
quantities were positive, and also $\mathcal{R}\geq1$.

\begin{acknowledgements}
B. Eslam Panah thanks University of Mazandaran.
\end{acknowledgements}

\end{document}